# Condensation droplet sieve

Chen Ma[1,2*], Zhiping Yuan[1,2], Li Chen[1,2], Lin Wang[1,2,4], Wei Tong[1,2,4], Cunjing Lv[1,2*]

& Quanshui Zheng[1,2,3*]

[1] Department of Engineering Mechanics, Tsinghua University, Beijing 100084, China

[2] Center for Nano and Micro Mechanics, Tsinghua University, Beijing 100084, China

[3] State Key Laboratory of Tribology, Tsinghua University, Beijing 100084, China

[4] Institute of Superlubricity Technology, Research Institute of Tsinghua University in Shenzhen, Shenzhen 518057, China

* Corresponding authors:

C. Ma: mac19@mails.tsinghua.edu.cn;

C. Lv: cunjinglv@mail.tsinghua.edu.cn;

Q. Zheng: zhengqs@mail.tsinghua.edu.cn;

## Abstract

Large droplets emerging during dropwise condensation impair surface properties such as anti-fogging/frosting ability and heat transfer efficiency. How to spontaneously detach massive randomly distributed droplets with controlled sizes has remained a great challenge. Herein, we present a general solution called condensation droplet sieve, through fabricating microscale thin-walled lattice (TWL) structures coated with a superhydrophobic layer. Growing droplets were observed to jumped off this TWL surface with 100% probability once becoming slightly larger than the lattices. The maximum radius and residual volume of droplets were strictly confined to 16 μm and 3.2 nl/mm$^2$ respectively, greatly surpassing the current state of the art. We reveal that this extremely efficient jumping is attributed to the large tolerance of coalescence mismatch and effective isolation of droplets between neighbouring lattices. Our work provides a new perspective for the design and fabrication of high-performance anti-dew materials.

# Introduction

Condensation droplets have been observed to self-propel and jump off superhydrophobic surfaces by virtue of the kinetic energy transferred from excessive surface energy released by coalescence[1]. Droplets with a typical radius of 10–100 μm have been reported to jump[2-5], way smaller than the capillary length (≈ 2.7 mm) required for gravity-induced shedding[6]. On account of this highly efficient removal of droplets before accumulation, superhydrophobic surfaces have been found advantageous in many fields including anti-fogging[7-9], anti-icing[10-12] and highly efficient heat transfer[13-15]. However, coalescence-induced jumping can fail because of the high adhesion of the nano-Wenzel state under high supersaturation[16-18], prominent viscous dissipation at small scale[19-21] and flow asymmetry with large coalescence mismatch[22-24]. It has been recently found that a nanocone surface inspired by nature with extremely low adhesion can significantly improve the jumping probability and reduced the maximum droplet radius to 35 μm[25]. Moreover, on a nanostructured CuO surface, with the external aid of a strong electric field, the maximum radius of charged condensation droplets was confined to 25 μm[26]. In addition, microscale structures and patterns have been shown to improve jumping by manipulating the nucleation and growth[27-29], introducing an out-of-plane Laplace pressure difference to extrude small droplets[30,31], enlarge the propelling force[32,33] or even trigger non-coalescence jumping[34]. However, the massive number and random distribution of condensation droplets make their coalescence and jumping behaviours hard to control. Further suppressing and realizing a strict control over droplet sizes still remain great challenges, but are essential for a scientific breakthrough. A delicate design of micro-structures that is able to make the chaotic distribution and jumping dynamics of droplets controllable is the key to tackle this problem. In this study, we fabricated a superhydrophobic surface with microscale thin-walled lattice (TWL) structures and achieved 100% jumping probability at a narrow range of droplet radius. On account of this, total removal of droplets slightly larger than the lattices by the TWL surface was identified without external aids such as wind[35], gravity[6] or electric field[26], similar to a condensation droplet sieve that is capable of filtering out all droplets larger than specific size.

# Results

## Thin-walled lattice structures

The TWL structures were fabricated on a silicon wafer using photolithography. Figure 1a shows an illustration and a scanning electron micrograph. The width of a single period, wall thickness and wall height were designed to be $W = 20$ μm, $D = 1$ μm, and $H = 10$ μm respectively. Structures were further coated with a layer of silanized silica nanobeads[30,34]. As shown in Fig. 1a, the thickness of the blurry layer is approximately 250 nm, which makes the walls slightly thicker than the design. The planar silicon sample decorated with such nanobeads shows superhydrophobicity with an equilibrium contact angle of $\theta_e = 165.1° \pm 0.7°$, an advancing contact angle of $\theta_a = 168.1° \pm 0.3°$ and a receding contact angle of $\theta_r = 161.8° \pm 1.3°$ (Supplementary Fig. S1).

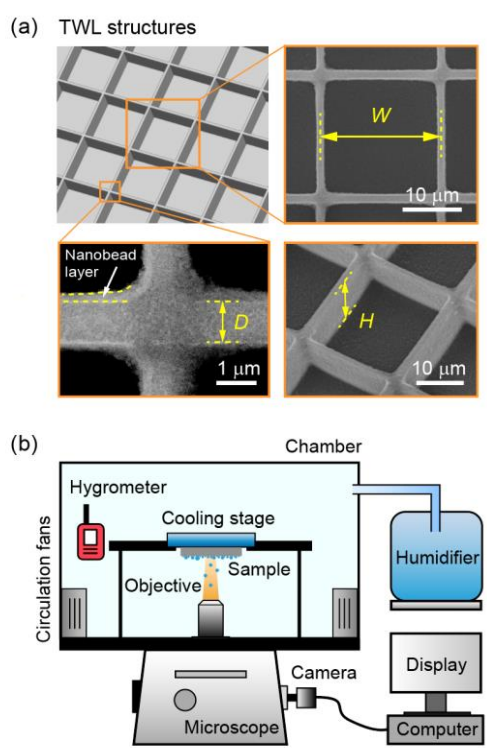

**Figure 1 | Surface characterization and experimental setup. a,** Illustration and scanning electronic micrograph (JSM-IT300, JEOL Japan) of the TWL structures. Width of a single lattice is $W = 20$ μm. The walls of lattices have a thickness of $D = 1$ μm, and a height of $H = 10$ μm. An approximately 250 nm thick superhydrophobic nanobead layer was prepared on the structure. The walls with coating are about 1.5 μm in thickness. **b,** Illustration of the experiments. An inverted microscope was used to observe the breath figures on samples. Humidity and temperature of air during condensation were well controlled using a chamber. Saturated air was generated by a humidifier and blown into the chamber. Circulation fans can generate a mild flow to achieve a uniform humidity. Hydrometer was placed near the sample to monitor the relative humidity and temperature of the air.

The experimental setup is illustrated in Fig. 1b. The samples used for condensation were adhered to a cooling stage. To avoid the re-deposition of departed droplets, the cooling stage was mounted upside down on an inverted microscope. A chamber was designed to maintain a stable humid environment. The temperatures of the air and the samples were controlled at 25 °C and 4 °C, respectively, during condensation. The relative humidity was measured to be 85 ± 5% using a hydrometer and maintained well. This value corresponds to a high supersaturation of $S = 3.3 ± 0.2$, which is defined as the ratio between the vapor pressure at room temperature and the saturated vapor pressure at sample temperature[7]. The breath figures were recorded using an industrial camera at a frame rate of 1 fps for 1 h.

**Distribution of droplet sizes**

We compared the breath figures of the planar surface and the TWL surface coated with Glaco layer after 1 h of condensation. In Fig. 2a, distinct differences appear between the two figures recorded with a visual field of 1.04 mm × 0.66 mm. The breath figure shows a chaotic pattern with a large range of droplet radii on the planar surface. In contrast, the breath figure on the TWL surface is regularly patterned without large droplets. The lower panels of Fig. 1a gives partially enlarged images. The positions of droplets on the planar surface are disordered while those on the TWL surface fall regularly in the lattices enclosed by thin walls, isolated from their neighbouring droplets. Comparison with a larger visual field (2.08 mm × 1.31 mm) is given in Supplementary Movie 1.

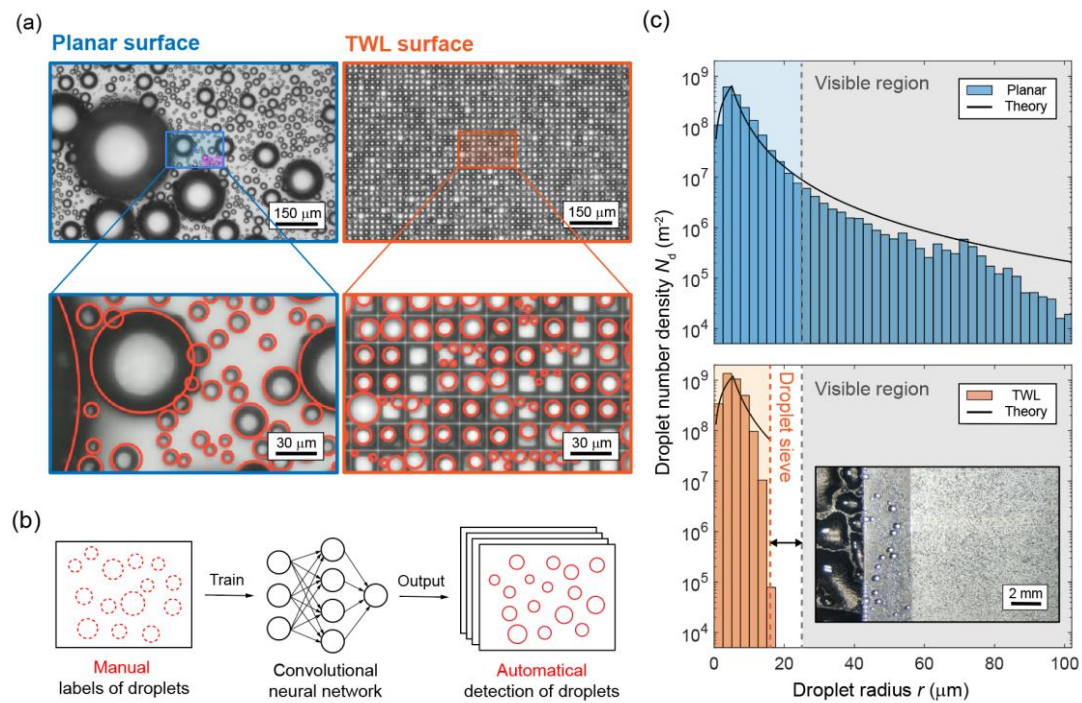

**Figure 2 | Breath figures on the planar and TWL surfaces. a,** Breath figures on both surfaces shot using a 10× objective. Magnified images of corresponding local breath figures are also given. Red circles are the detection results made by our convolutional neural network (CNN). **b,** Illustration of the construction of the CNN. Manually made labels are used to train the CNN. After training and adjustments, the CNN can help detect a series of images automatically. **c,** Distribution of the droplet radii. Bars represent the number density of droplets in evenly spaced radius ranges 2.5 μm in width. Blue bars in the upper panel represent the distribution on the planar surface while orange ones in the low panel represent that on the TWL surface. Gray regions represent the radius range that is visible to naked eyes. Solid black curves show the theoretical distributions. Three surfaces from the left to the right in the chart of the TWL surface demonstrates the breath figure at a centimeter scale, corresponding the cooling stage surface, planar surface and TWL surface, respectively.

To extract quantified information from the breath figures, we need to know the positions and radii of thousands of droplets in vision. However, the number of frames recorded during 1 h is huge. It is also difficult to recognize droplets accurately using traditional image processing methods[36] on the TWL surface because vacant lattices could be misrecognized as droplets. Recently, deep learning have shown its potential in accurate droplet detection[37,38]. Consequently, we developed a convolutional neural network (CNN) as shown in Fig. 2b (details are provided in Supplementary Discussion 2). Only a small number of manual labels are needed to train the CNN. After repeated training and adjustments, droplets can be accurately discerned in a series of breath figures (Supplementary Movie 2). The results are shown in Fig. 2a as the red circles. Droplets partially hiding below the periphery of large droplets and those sitting at the corners of the lattices were detected successfully.

Based on the CNN results, the droplet number density $N_d$ (number of droplets per projected area) in evenly spaced droplet radius ranges from 0 to 100 μm is shown in Fig. 2c. The distribution of $N_d$ is averaged by 2400 frames of breath figures after the condensation reaches a steady state at 1200 s. On the planar surface, the droplet radii are loosely distributed. Many of them fall in the region colored gray, where droplets are large enough to be visible to the naked eye with a limit of 50 μm in diameter, corresponding to an observing distance of 20 cm[39]. The maximum radius $r_M$ observed in our experiment is 173 μm. On the contrary, droplet sizes are distributed intensively in a narrow range of droplet radii on the TWL surface, with a maximum radius of 16 μm, smaller than the limit of human vision. This result reveals that there exists a condensation droplet sieve that filters out all large droplets, denoted using the orange dashed line. The black solid curves are the theoretical distributions, which are highly consistent with the experimental results (details for the theory in Supplementary Discussion 3).

The inset in the chart of the TWL surface shows the centimeter-scale breath figures on the bare cooling stage surface, planar surface and TWL surface from the left to the right. Puddles that are several millimeters large form on the metallic surface of the cooling stage. Visible droplets with radii at the scale of 100 μm remain on the planar surface. In comparison, the breath figure on the TWL surface is uniform without visible sessile droplets. Even though dark spots that represent droplets at the scale of 10 μm are observable in this macro photo, they are not detectable by the naked eye when observed from a distance of 20 cm.

**Mechanism of condensation droplet sieve**

The dynamic process of coalescence was investigated. The left panels of Fig. 3a is the time-lapse of droplets on the planar surface (Supplementary Movie 3). The visual field is 24 μm in height. Droplets about to coalesce are outlined by yellow dashed lines. Researchers have previously found that the flow asymmetry[22,23] and a weak reaction force of surface towards droplet bridge[24] could prevent jumping under large coalescence mismatch such as the case of droplets A and B. Moreover, the growing effect of adhesion and viscous dissipation with decreasing droplet sizes has been found to hold back jumping for droplets smaller than 10 μm[3,19,20,40], which accounts for the failure of jumping for droplets C and D with a radius of 6 μm. The number of droplets that participate in coalescence in different radius ranges were counted as shown in the chart. Dark and light blue bars represent the number of droplets that jump or do not jump after coalescence of totaling 500 successive coalescences within an condensation area of 0.38 mm × 0.21 mm. The solid line with circular dots is the jumping probability which is calculated as the proportion of jumping droplets after coalescence[7,41]. For small droplets, the jumping probability is almost zero. However, it reaches a level of approximately 35% for relatively large droplets.

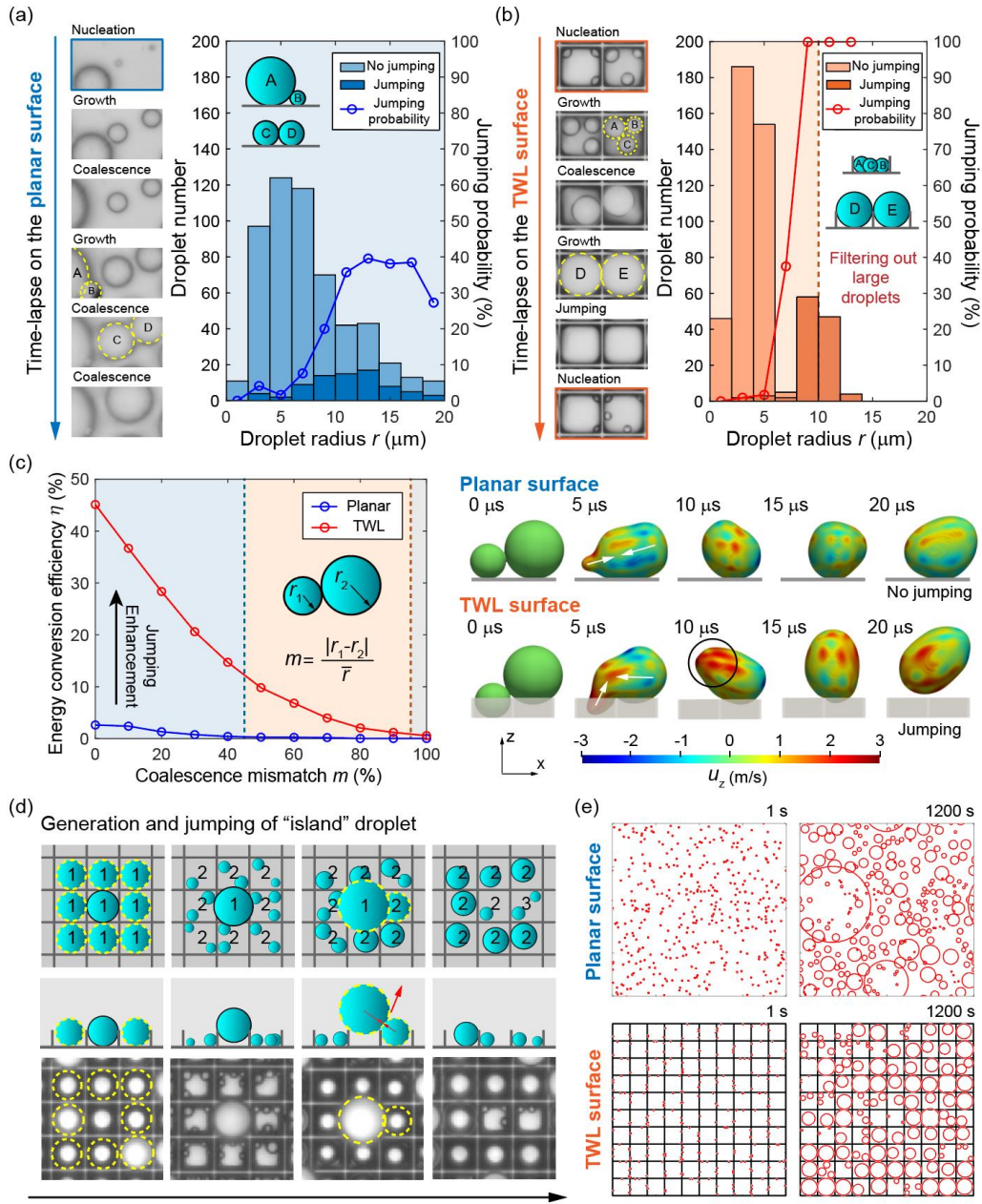

**Figure 3 | Investigation and verification of the mechanisms. a,** Time lapse and chart of jumping probability on the planar surface. Insets in the chart show the droplets corresponding to those in the time-lapse. Dark and light bars give the number of droplets that jump or do not jump after coalescence respectively in evenly distributed radius ranges of 2 μm. The Solid line with circular data dots is the jumping probability. **b,** The arrangement here is the same as Fig. 3a, but the results are from the TWL surface. **c,** Jumping dynamics on the two surfaces from VOF simulations. The left panel is the chart showing variation of energy-conversion efficiency with coalescence mismatch. The right panel is the time lapse of droplet coalescence from VOF simulations. **d,** Sketches and experimental observations of the generation and jumping of "island" droplet. **e,** Simulation of overall condensation process on the two surfaces. Droplets are represented by red circles. Small droplets could hide underneath large droplets. Black lines in the simulation on the TWL surface denote the thin walls.

The area of the lattice top is small compared with the actual surface area of the TWL surface (< 0.1), which makes the droplets tend to nucleate at the bottom or the side walls of the lattices. As shown in Fig. 3b, the droplets first nucleate inside the lattices, followed by growth and coalescence (Supplementary Movie 3). Due to the small size of the droplets and the confinement of lattice walls, droplets A, B and C do not jump but merge into a large one. This leads to a nearly zero jumping probability for small droplets inside lattices. Nevertheless, as the individual droplet grows large enough to touch its neighbouring droplets, they jump off and leave the blank lattice for new nucleation. This corresponds to the abrupt increase in the jumping probability to 100% when droplets radii are comparable to half the lattice width (10 μm). Consequently, all large droplets are filtered out by this complete removal. It should be noted that droplets fall in the range of 6–8 μm hardly coalesce because they are too large to leave room for the generation of another condensate droplet in the lattice and too small to touch the neighbouring droplets due to the thin wall enclosure.

The existence of the thin walls between the lattices are the main reason for this highly efficient jumping. The left panel of Fig. 3c shows the relationship between the energy-conversion efficiency and the coalescence mismatch after two-droplet coalescence (Supplementary Movie 4). The result is calculated using the customized VOF solver JumpingFOAM based on OpenFOAM, which is particularly designed to handle jumping droplet simulations[42,43] (Supplementary Discussion 4). The energy-conversion efficiency $\eta$ is defined as the ratio of translational kinetic energy of jumping droplets to the surface energy released after coalescences[3,24,40]. Different manners exist to quantify coalescence mismatch $m$[22-24]. Here, we define it as the ratio of the radius difference to the average radius. The droplet radii on the two surfaces were controlled to be the same for different mismatches. On the TWL surface, $\eta$ is enhanced by an order of magnitude compared to the planar surface. In the blue region, with a mismatch of < 45% (corresponding to a size ratio of 1:1.58), droplets jump on both surfaces. However, when the mismatch rises into the orange region, only the TWL surface can still ensure successful jumping until a mismatch of 95% (corresponding to a size ratio of 1:2.64).

Previous work made on macroscale droplets has shown that jumping is significantly enhanced on a superrepellent surface with a ridge placed between two droplets[44]. This is due to the effective redirection of in-plane velocity vectors to out-of-plane velocity vectors by the ridge[44]. Our thin walls serve the same role for the microscale coalescences. On the right panel of Fig. 3c is the detailed behaviour of droplets coalescing on the planar and TWL surfaces with $m$ = 50%. The droplet surfaces are colored with an off-plane velocity, $u_z$. At the initial moment of 5 μs on the planar surface, liquid from two drops flows in countering directions and crashes into each other as depicted by the white arrows, followed by *in situ* oscillations without jumping. On the TWL structure, with the block of thin walls, liquid is induced to flow in the off-plane direction. Therefore, a region of the droplet obtains a high $u_z$ as labeled by a circle. The strong off-plane momentum helps to overcome the liquid-solid adhesion. In addition, the easy detachment also benefits from the low adhesion of large droplet suspended in the Cassie state.

The time lapses in Fig. 3d illustrate the generation of the largest coalescence mismatch on the TWL surface. Considering that nine droplets grew large enough to coalesce with the neighbouring droplets as shown in the first column. Let us denote them as the 1st-generation droplets. Unfortunately, all the droplets, except the one in the center, jump off, leaving the central droplet as an "island". It continues to grow without coalescence, while the 2nd-generation droplets nucleate and grow simultaneously (the second column). When the nearest 2nd-generation droplet touches the 1st-generation "island", the largest coalescence mismatch occurs as shown in the third column. The mismatch measured in this experimental figure is 49%, which is still within the range of the TWL surface to trigger jumping (95%, see Fig. 3c). More importantly, thanks to the enclosure by the thin walls, the 2nd-generation droplets are isolated from neighbouring droplets except for the large "island". This isolation effect ensures that at least one 2nd-generation droplets will ultimately touch the "island". In other words, the "island" will coalesce within the growth time of 2 generations (the fourth column), which in turn guarantees a small coalescence mismatch and its successful jumping.

To further validate the above mechanism, we customized a program based on recent works[45,46] to simulate the overall condensation process (Supplementary Fig. S3 and Movie 5). The simulation box is 200 μm in both width and length with periodic boundary conditions (details are provided in Supplementary Discussion 5). The growth curves of droplets on two surfaces were measured from experiments (Fig. S4). The initial nucleation sites were randomly placed, as shown in Fig. 3e at the first second. When the simulation was run for 1200 s, large droplets emerged disorderedly on the planar surface. On the contrary, droplets are arranged regularly without large droplets on the TWL surface, which is in good accordance with experiments.

**Variation of condensation properties with time**

The maximum radius and the residual volume of droplets are two important parameters to characterize the anti-dew ability[7,45,47]. We monitored their variation within 1 h of condensation. The condensation time starts from the cooling of the samples. Figure 4b shows how the maximum radius $r_M$ varies with time. On the planar surface, $r_M$ keeps growing almost linearly with time, until it reaches a peak of 173 μm at $t$ = 3407 s (moment ① on the planar surface), followed by a series of sudden falls ending at $t$ = 3514 s (moment ② on the planar surface). The two breath figures on the left of Fig. 4a correspond to the aforementioned two moments, showing that jumping of several large droplets leads to the dramatic falls. As mentioned above, the coalescence mismatch significantly deteriorates energy-conversion efficiency. Thus, if a large droplet is formed, it can hardly jump by coalescing with the neighbouring droplets. These coalescences serve no use but further feed the droplets making it larger and even harder to detach. These vicious circles continue until the large droplet touches others comparable to its size. In comparison, $r_M$ reaches a steady value after 1200 s and is well controlled below 16 μm with a time-averaged value of only 13.8 μm on the TWL surface. This result is even smaller than the state of the art using delicate nanocone surface with a maximum radius of 35 μm[25], which is denoted by the black dashed line in Fig. 4b.

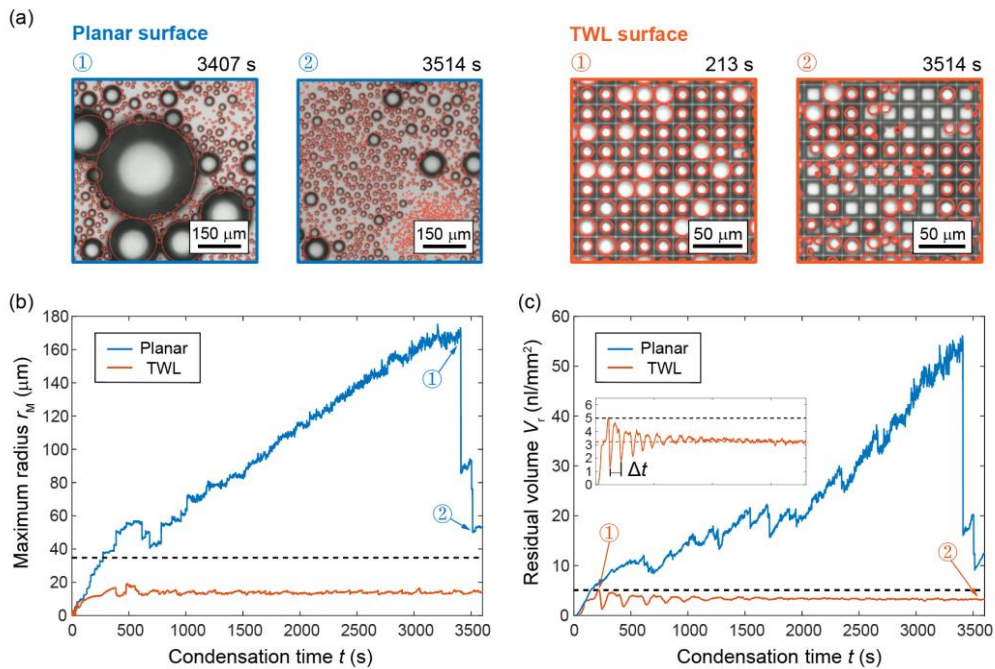

**Figure 4 | Variation of condensation properties with time. a,** Breath figures at certain moments during condensation. The left-hand side shows two figures on the planar surface. The condensation time for moment ① is 3407 s while that for

moment ② is 3514 s. The right-hand side shows two figures on the TWL surface. The condensation time for moment ① is 213 s while that for moment ② is 3514 s. **b,** Variation of maximum radius with time on two surfaces. Points ① and ② correspond to the breath figures on the planar surface in Fig. 4a. **c,** Variation of residual volume with time on two surfaces. Points ① and ② correspond to the breath figures on the TWL surface in Fig. 4a. Inset shows the enlarged curve of residual volume on the TWL surface. The black dashed lines in **b** and **c** represent the state-of-the-art results[7,25].

Figure 4c shows the time evolution of the residual volume $V_r$ defined as the total volume of remaining droplets per area with a unit of nl/mm$^2$. Because the volume of a droplet is related to the third power of its radius, the residual volume is mainly determined by the volume of large droplets. Consequently, the variation tendency of $V_r$ is similar to that of $r_M$. On the TWL surface, $V_r$ reaches a stable value of 3.2 nl/mm$^2$, which is also smaller than the state-of-the-art result (5 nl/mm$^2$)[7]. The enlarged curve on the TWL surface in the inset shows a trend similar to damped oscillation with a period of $\Delta t$ = 192 s, which approximates the time span required for the refreshing of a single lattice. The peak for this curve is reached at moment ① on the TWL surface. As shown in the corresponding breath figure on the right of Fig. 4a, most cells are filled with sole and large droplets with a small variation in size, which increases the residual volume. The uniformity of droplet sizes at moment ① is owing to the simultaneous nucleation at the beginning of condensation. In contrast, at moment ②, when each lattice has experienced numerous refreshes, the corresponding breath figure shows a scattered distribution with few large droplets, which makes the residual volume at a smaller level. This phenomenon is also predicted by simulation as shown in Supplementary Fig. S3.

**Design criterion for condensation droplet sieve**

The influence of lattice geometries on the droplet removal efficiency is further discussed. We carried out experiments on three other TWL surfaces having various structure shapes and sizes. As shown in Fig. 5a, S and H represent the square and hexagonal lattices, respectively. The number that follows S or H shows the width of the lattice $W$ in the unit of micrometers. The width of the hexagonal lattices is defined as the distance between the opposite thin walls. The thickness $D$ of all these TWL surfaces is 1 μm, while the lattice height $H$ maintains as half of $W$. TWL S-20 is the original TWL surface that we discussed in the above, while the three other TWL surfaces have different geometries.

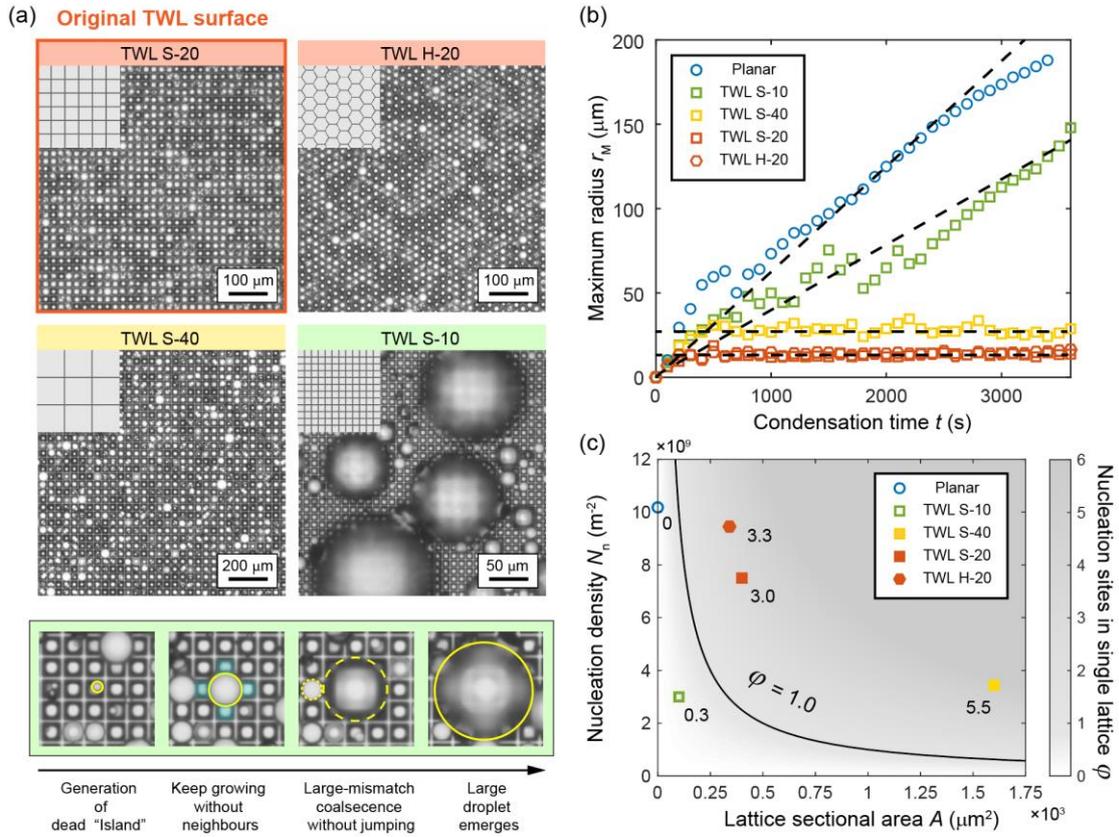

**Figure 5 | The criterion for designing functional condensation droplet sieve. a,** Breath figures on different TWL surfaces after 1 h of condensation (upper panels) and time lapse of droplet coalescence on TWL S-10 (lower panels). The geometrical parameters for the 4 surfaces are as follows: TWL S-20, $W$ = 20 μm, $H$ = 10 μm, $D$ = 1 μm; TWL H-20, $W$ = 20 μm, $H$ = 10 μm, $D$ = 1 μm; TWL S-40, $W$ = 40 μm, $H$ = 20 μm, $D$ = 1 μm; TWL S-10, $W$ = 10 μm, $H$ = 50 μm, $D$ = 1 μm. **b,** Variation of maximum droplet radius $r_M$ with condensation time $t$ on different surfaces. Dashed solid lines indicate their variation tendency. **c,** Color map showing the relationship between $\varphi$, $N_n$ and $A$. The black line is the critical line separating functional (solid dots) and unfunctional condensation droplet sieves (hollow dots). The numbers accompanying the dots shows their values of $\varphi$.

The mechanism of condensation droplet sieve is not restricted to the shape of the lattice. This is validated as shown in the upper two breath figures in Fig. 5a. No matter on the "chessboard" TWL S-20 or the "honeycomb" TWL H-20, the condensation droplet sieves are obtained. We further considered whether the sieve maintains on the TWL surfaces with smaller or larger lattice sizes. The lower two breath figures correspond to TWL structures that are twice and half the size of the original TWL surface, respectively. For the magnified structures (TWL S-40), droplet sizes are well confined just like the original one. However, on minified structures (TWL S-10), droplets show disordered distribution with large droplets, indicating the failure of condensation droplet sieve. This failure is further verified by the evolution of maximum droplet radius with time in Fig. 5b. The maximum radius quickly reaches a stable level on surfaces TWL H-20, TWL S-20 and TWL S-40. Moreover, the stable maximum radius is almost proportional to the lattice width. However, it does not mean droplet sizes can be further decreased by shrinking the lattices unlimitedly. On TWL S-10, the maximum droplet radius increases out of control just like the planar surface (Supplementary Movie 6), which indicates that there seems to be a critical size of the lattice to realize the condensation droplet sieve.

In the low panel of Fig. 5a, we look into how the small-sized TWL surface fails. At the beginning of the time lapse, nucleation site emerges in the middle lattice. However, as it grows larger, the neighbouring cells are still vacant, which makes the central droplet an "island" without neighbouring droplet. If a cell is surrounded by such nucleation-free lattices, the droplet in it becomes a dead "island". Different from the transient "island" on the original TWL surface, the dead "island" here is permanent because nucleation-free cells are dried out without further supplement of droplets. As a result, the "island" droplet will have to grow to at least 3-lattice-wide to coalesce with others. However, the coalescence mismatch at that point is too large, which is about 103% as shown in Fig. 5a, larger than the critical mismatch for jumping on the TWL surface. Consequently, the droplet grows under the feeding of small droplets and eventually becomes a large droplet.

Above observation indicates that there exists a design criterion between nucleation density (number of nucleation site per projected area) and lattice sectional area to avoid dead "island" droplets. The average number of nucleation sites in the single lattice $\varphi$ should be larger than unit so that lattices will have a large possibility to be filled with nucleation sites. The nucleation density $N_n$ and lattice sectional area $A$ together determine $\varphi$ with a relationship of $\varphi = N_n A$. Consequently, we get the criterion of realizing condensation droplet sieve which writes: $N_n A > 1$. This criterion is well verified as depicted in Fig. 5c. Background color shows the distribution of $\varphi$ and the solid line represents the critical situation where $\varphi = 1.0$. Nucleation density of each surface is obtained from their breath figures. Planar surface can also be drawn in this color map, if we regard it as a surface with infinitely small lattices. The critical line successfully separates the solid spots that represent surfaces able to sieve droplets and hollow ones show uncontrollable droplet growth. This criterion shows that with a given $N_n$, $A$ has a minimum value, and the value increases with decreasing $N_n$. In real word applications, $N_n$ could vary with humidity, substrate temperature[48] and the types of superhydrophobic coatings[49]. Thus, it is necessary to assess the smallest $N_n$ in varying condensation conditions to determine the design of functionable condensation droplet sieve.

## Discussion

Condensation droplet sieve is, to the best of our knowledge, realized for the first time with strict confinement of the maximum radius and residual volume on the TWL surfaces. Jumping enhancement provided by thin walls increases the tolerance of the coalescence mismatch. Thin walls also isolate the droplets from neighbouring ones to ensure the jumping of the "island" droplets before they grow too large. These two mechanisms give a supreme jumping efficiency with 100% jumping probability. This newly proposed strategy reveals that surface with high jumping efficiency can be realized on microscale structures without delicate design of hard-to-control nanostructures. We also proposed a design criterion related to nucleation density and the scale of lattices to ensure the functionality of condensation droplet sieves. Future works are needed to fully exploit the potentials of this new concept in anti-frosting and highly efficient heat transfer. To conclude, our work demonstrates a simple and novel method for high-performance anti-dew materials which have promising applications in numerous fields[7,12,15].

# References

1    Boreyko, J. B. & Chen, C.-H. Self-propelled dropwise condensate on superhydrophobic surfaces. *Phys. Rev. Lett.* **103**, 184501 (2009).

2    Enright, R. *et al.* How coalescing droplets jump. *ACS Nano* **8**, 10352-10362 (2014).

3    Liu, F., Ghigliotti, G., Feng, J. J. & Chen, C.-H. Numerical simulations of self-propelled jumping upon drop coalescence on non-wetting surfaces. *J. Fluid Mech.* **752**, 39-65 (2014).

4    Nam, Y., Kim, H. & Shin, S. Energy and hydrodynamic analyses of coalescence-induced jumping droplets. *Appl. Phys. Lett.* **103**, 161601 (2013).

5    Nam, Y., Seo, D., Lee, C. & Shin, S. Droplet coalescence on water repellant surfaces. *Soft Matter* **11**, 154-160 (2015).

6    Tanasawa, I. in *Proceedings of 5th International Transfer Conference.* 188 (1974).

7    Mouterde, T. *et al.* Antifogging abilities of model nanotextures. *Nat. Mater.* **16**, 658-663 (2017).

8    Sun, Z. *et al.* Fly-eye inspired superhydrophobic anti-fogging inorganic nanostructures. *Small* **10**, 3001-3006 (2014).

9    Wang, Q., Yao, X., Liu, H., Quéré, D. & Jiang, L. Self-removal of condensed water on the legs of water striders. *Proc. Natl. Acad. Sci. U.S.A.* **112**, 9247-9252 (2015).

10   Zhang, Q. *et al.* Anti-icing surfaces based on enhanced self-propelled jumping of condensed water microdroplets. *ChemComm* **49**, 4516-4518 (2013).

11   Boreyko, J. B. & Collier, C. P. Delayed frost growth on jumping-drop superhydrophobic surfaces. *ACS Nano* **7**, 1618-1627 (2013).

12   Lv, J., Song, Y., Jiang, L. & Wang, J. Bio-inspired strategies for anti-icing. *ACS Nano* **8**, 3152-3169 (2014).

13   Miljkovic, N. *et al.* Jumping-droplet-enhanced condensation on scalable superhydrophobic nanostructured surfaces. *Nano Lett.* **13**, 179-187 (2013).

14   Wen, R., Xu, S., Ma, X., Lee, Y.-C. & Yang, R. Three-dimensional superhydrophobic nanowire networks for enhancing condensation heat transfer. *Joule* **2**, 269-279 (2018).

15   Cho, H. J., Preston, D. J., Zhu, Y. & Wang, E. N. Nanoengineered materials for liquid–vapour phase-change heat transfer. *Nat. Rev. Mater.* **2**, 1-17 (2016).

16   Wen, R. *et al.* Wetting transition of condensed droplets on nanostructured superhydrophobic surfaces: coordination of surface properties and condensing conditions. *ACS Appl. Mater. Interfaces* **9**, 13770-13777 (2017).

17   Wen, R. *et al.* Hydrophobic copper nanowires for enhancing condensation heat transfer. *Nano Energy* **33**, 177-183 (2017).

18   Feng, J., Pang, Y., Qin, Z., Ma, R. & Yao, S. Why condensate drops can spontaneously move away on some superhydrophobic surfaces but not on others. *ACS Appl. Mater. Interfaces* **4**, 6618-6625 (2012).

19   Wang, F.-C., Yang, F. & Zhao, Y.-P. Size effect on the coalescence-induced self-propelled droplet. *Appl. Phys. Lett.* **98**, 053112 (2011).

20   Lv, C. *et al.* Condensation and jumping relay of droplets on lotus leaf. *Appl. Phys. Lett.* **103**, 021601 (2013).

21   Lecointre, P. *et al.* Ballistics of self-jumping microdroplets. *Phys. Rev. Fluid* **4**, 013601 (2019).

22   Wasserfall, J., Figueiredo, P., Kneer, R., Rohlfs, W. & Pischke, P. Coalescence-induced droplet jumping on superhydrophobic surfaces: Effects of droplet mismatch. *Phys. Rev. Fluid* **2**, 123601

(2017).

23   Wang, K. *et al.* Critical size ratio for coalescence-induced droplet jumping on superhydrophobic surfaces. *Appl. Phys. Lett.* **111**, 061603 (2017).

24   Yan, X. *et al.* Droplet jumping: effects of droplet size, surface structure, pinning, and liquid properties. *ACS Nano* **13**, 1309-1323 (2019).

25   Lecointre, P. *et al.* Unique and universal dew-repellency of nanocones. *Nat. Commun.* **12**, 1-9 (2021).

26   Miljkovic, N., Preston, D. J., Enright, R. & Wang, E. N. Electric-field-enhanced condensation on superhydrophobic nanostructured surfaces. *ACS Nano* **7**, 11043-11054 (2013).

27   Han, T., Kwak, H. J., Kim, J. H., Kwon, J.-T. & Kim, M. H. Nanograssed zigzag structures to promote coalescence-induced droplet jumping. *Langmuir* **35**, 9093-9099 (2019).

28   Hou, Y., Yu, M., Chen, X., Wang, Z. & Yao, S. Recurrent filmwise and dropwise condensation on a beetle mimetic surface. *ACS Nano* **9**, 71-81 (2015).

29   Xing, D., Wang, R., Wu, F. & Gao, X. Confined growth and controlled coalescence/self-removal of condensate microdrops on a spatially heterogeneously patterned superhydrophilic–superhydrophobic surface. *ACS Appl. Mater. Interfaces* **12**, 29946-29952 (2020).

30   Lv, C., Hao, P., Zhang, X. & He, F. Dewetting transitions of dropwise condensation on nanotexture-enhanced superhydrophobic surfaces. *ACS Nano* **9**, 12311-12319 (2015).

31   Wen, R. *et al.* Hierarchical superhydrophobic surfaces with micropatterned nanowire arrays for high-efficiency jumping droplet condensation. *ACS Appl. Mater. Interfaces* **9**, 44911-44921 (2017).

32   He, M., Ding, Y., Chen, J. & Song, Y. Spontaneous uphill movement and self-removal of condensates on hierarchical tower-like arrays. *ACS Nano* **10**, 9456-9462 (2016).

33   Tang, Y., Yang, X., Li, Y., Lu, Y. & Zhu, D. Robust micro-nanostructured superhydrophobic surfaces for long-term dropwise condensation. *Nano Lett.* **21**, 9824-9833 (2021).

34   Yan, X. *et al.* Laplace pressure driven single-droplet jumping on structured surfaces. *ACS Nano* **14**, 12796-12809 (2020).

35   Torresin, D., Tiwari, M. K., Del Col, D. & Poulikakos, D. Flow condensation on copper-based nanotextured superhydrophobic surfaces. *Langmuir* **29**, 840-848 (2013).

36   Martin, H., Barati, S. B., Pinoli, J.-C., Valette, S. & Gavet, Y. in *7th International Conference on Computer Science, Engineering & Applications (ICCSEA 2017).* 115 à 126 (AIRCC Publishing Corporation, 2017).

37   Suh, Y. *et al.* A deep learning perspective on dropwise condensation. *Adv. Sci.*, 2101794 (2021).

38   Yan, J., Ma, R. & Du, X. Consistent optical surface inspection based on open environment droplet size-controlled condensation figures. *Meas. Sci. Technol* (2021).

39   Yanoff, M., Duker, J. S. & Augsburger, J. J. *Ophthalmology*.  (2020).

40   Cha, H. *et al.* Coalescence-induced nanodroplet jumping. *Phys. Rev. Fluid* **1**, 064102 (2016).

41   Rykaczewski, K. *et al.* Multimode multidrop serial coalescence effects during condensation on hierarchical superhydrophobic surfaces. *Langmuir* **29**, 881-891 (2013).

42   Yuan, Z., Hou, H., Dai, L., Wu, X. & Tryggvason, G. Controlling the jumping angle of coalescing droplets using surface structures. *ACS Appl. Mater. Interfaces* **12**, 52221-52228 (2020).

43   Yuan, Z. *et al.* Ultimate jumping of coalesced droplets on superhydrophobic surfaces. *J. Colloid Interface Sci.* **587**, 429-436 (2021).

44   Vahabi, H., Wang, W., Mabry, J. M. & Kota, A. K. Coalescence-induced jumping of droplets on superomniphobic surfaces with macrotexture. *Sci. Adv.* **4**, eaau3488 (2018).

45   Birbarah, P., Chavan, S. & Miljkovic, N. Numerical simulation of jumping droplet condensation. *Langmuir* **35**, 10309-10321 (2019).

46   Hu, Z., Yuan, Z., Hou, H., Chu, F. & Wu, X. Event-driven simulation of multi-scale dropwise condensation. *Int. J. Heat Mass Transf.* **167**, 120819 (2021).

47   Mukherjee, R., Berrier, A. S., Murphy, K. R., Vieitez, J. R. & Boreyko, J. B. How surface orientation affects jumping-droplet condensation. *Joule* **3**, 1360-1376 (2019).

48   Liu, X. & Cheng, P. Dropwise condensation theory revisited Part II. Droplet nucleation density and condensation heat flux. *Int. J. Heat Mass Transf.* **83**, 842-849 (2015).

49   Aili, A., Ge, Q. & Zhang, T. How nanostructures affect water droplet nucleation on superhydrophobic surfaces. *J. Heat Transfer* **139** (2017).